\documentclass[aps,12pt]{revtex4}
\usepackage[dvips]{epsfig}

\begin{document}

\title{The Tails of  the crossing probability}

\author{Oleg A.Vasilyev$^{1,2}$}

\email{vasilyev@itp.ac.ru}

\affiliation{%
$^1$ Laboratoire de Physique Th\'eorique de la Mati\`ere Condens\'ee, 
Universit\'e
Paris-VI, 75252 Paris Cedex 05, France \\ 
$^2$L.D.Landau Institute for Theoretical Physics RAS,
142432, Chernogolovka, Moscow reg., Russia}

\date{\today}

\bigskip

\begin{abstract}

The scaling of the tails of the probability of a system
to percolate only in the horizontal direction $\pi_{hs}$ 
was investigated numerically
 for correlated site-bond percolation model for $q=1,2,3,4$.
We have to demonstrate that the tails of the crossing probability 
far from the critical point $p_{c}$ have shape 
$\pi_{hs}(p) \simeq D\exp(c L\left[p-p_{c}\right]^{\nu})$ where 
$\nu$ is the correlation length index, $p=1-\exp(-\beta)$ is the
probability of a bond to be closed. At criticality we observe crossover
to another scaling $ \pi_{hs}(p) \simeq A 
\exp \left(-b \left\{ L [p-p_{c}]^{\nu}  \right\}^{z} \right)$.
Here $z$ is a scaling index
describing the central part of the crossing probability. 

\end{abstract}
\pacs{64.60.Ak, 05.10.Ln, 05.70.Jk}

\maketitle

\section{Introduction}

The $q$-state Potts model can be represented as the
correlated site-bond percolation in terms of Fortuin-Kasteleyn
clusters~\cite{FK}. At the critical point of the second order phase 
transition, the infinite cluster is formed.
This cluster crosses the system connecting the opposite sides of the square 
lattice. In the last decade the study of the shape of the crossing probability
was performed  by conformal  methods~\cite{Cardy0,Cardy1,Watts,Sm1,Sm2,Cardy2}
as well as  numerically~\cite{Lang1,Lang2,Ziff1,Hu1,Hu2,Hu3,Hu4,Lang3,NZ1,NZ2}.
 
According to Refs.~\cite{LSS,Wester} the distribution function
of the percolation thresholds is Gaussian function. 
Following the number of works~\cite{Ziff1,BW,HA,NZ1,NZ2} the 
tails of the distribution function are not Gaussian ones.
The authors of the recent work~Ref.~\cite{ONS}
are still uncertain to distinguish a stretched exponential 
behavior from a Gaussian.

\begin{figure}[h]
\epsfig{file=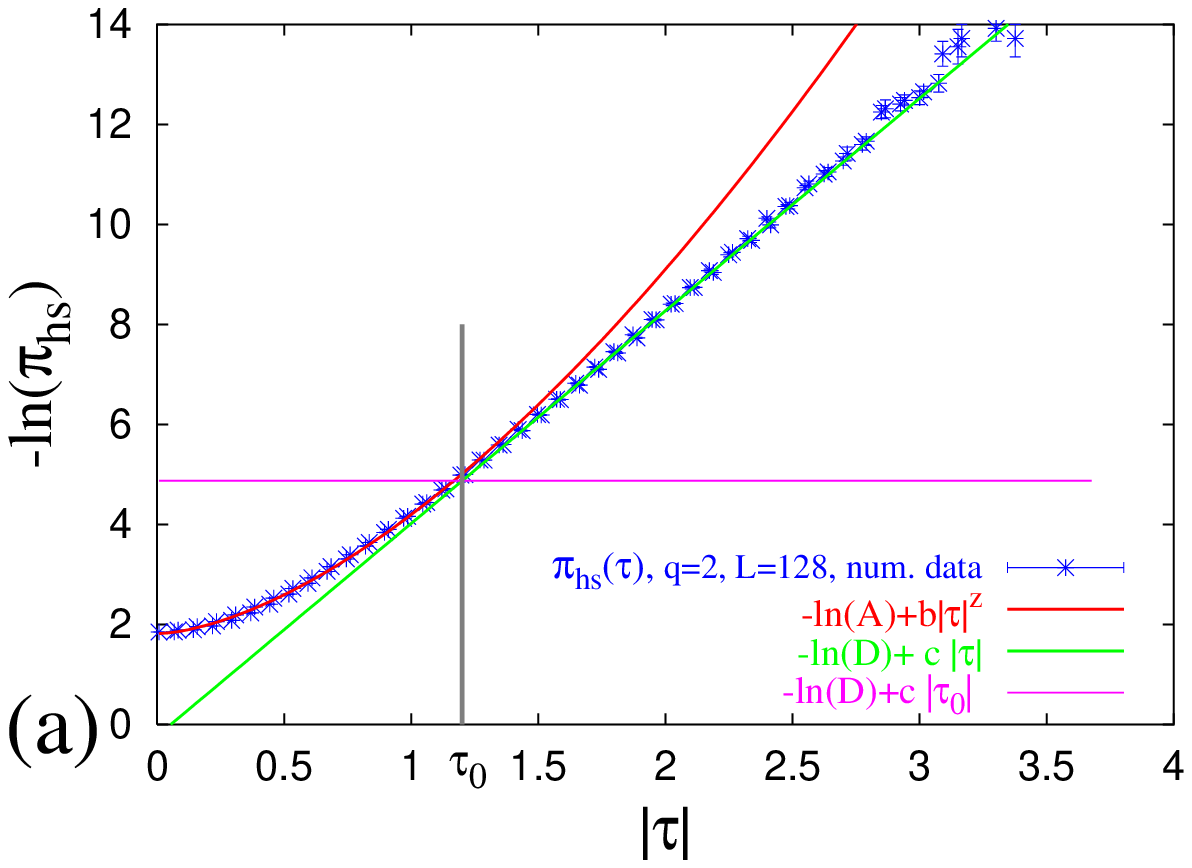,width=8.6cm,height=7cm}
\epsfig{file=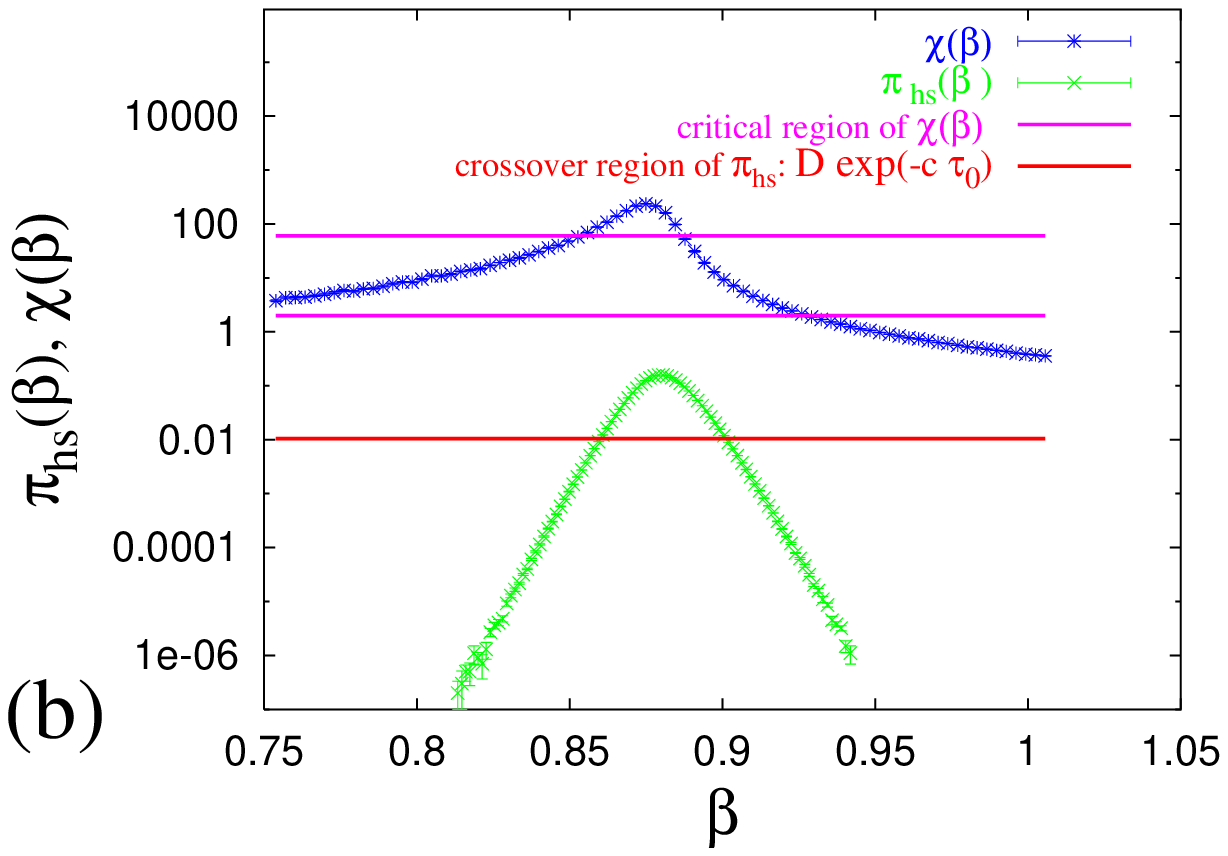,width=8.6cm,height=7cm}
\caption{
a) The absolute value of the logarithm of the crossing probability 
$-\ln(\pi_{hs})$ for the Ising model as a 
function of the absolute value of the scaling 
variable $\tau=L(p-p_{c})^{\nu}$ on the lattice $L=128$;
b) The magnetic susceptibility $\chi(\beta)$ and the crossing 
probability $\pi_{h}(\beta)$ as a functions of the 
inverse temperature $\beta$ for the Ising model, $L=128$.}
\label{fig1}
\end{figure}

The aim of this paper is to investigate
the shape of the probability of a system to percolate only in
horizontal direction $\pi_{hs}$.  We perform numerical simulation of  
correlated site-bond percolation model for $q=1,2,3,4$
(the percolation model $q=1$, the Ising model $q=2$ and the Potts 
model $q=3,4$) for lattice sizes $L=32,48,64,80,128$.
The scaling formulas for a body of the crossing probability at criticality
and for tails of the crossing probability were obtained. The final result
for the representative case $q=2$, $L=128$ is immediately presented in 
Fig.~\ref{fig1}a). 
Details of fitting procedure are described in Section~\ref{secapp2}.
In this figure we plot (by  crosses)
the numerical data for the absolute value of the logarithm of crossing 
probability $\pi_{hs}$ for the Ising model ($q=2$) on the lattice $L=128$
as a function of the absolute value of the scaling variable
$\tau=L(p-p_{c})^{\nu}$. Here $p=1-\exp(-\beta)$
is a probability of a bond to be closed, $\beta$ is the inverse 
temperature, $\nu$ is the correlation length scaling index.
 The critical point in the $p$ scale for the $q$-state Potts model 
is $p_{c}=\sqrt{q}/(1+\sqrt{q})$ see Ref.~\cite{Baxter}
and we get $p_{c}(q=2)\simeq 0.58578\dots$.

We can see from Fig.~\ref{fig1}a) that the function $\pi_{hs}(\tau)$
consists of two parts: the body $|\tau|<\tau_{0}$ and the tails 
$|\tau|>\tau_{0}$. The negative logarithm of the body of the crossing probability 
as a function of $\tau$ is well described by function $-\ln(A)+b |\tau|^{z}$ (solid line 
on the Fig.~\ref{fig1}a)). Here $z$ is some scaling index.  The value of the 
crossing probability at the critical point $\pi_{hs}(p_{c})=A$ may be 
computed (at least for percolation) by conformal field methods~\cite{Watts}. 
The negative logarithm of the tails of the crossing probability have 
shape $-\ln(D)+c \tau$ 
(dashed line in the Fig.~\ref{fig1}a)). This line is tangent
to the body at the point $\tau_{0}$. This point is marked by the horizontal  line.
Let us note that in Fig.~\ref{fig1}a) we plot two branches of the crossing
probability (for $\tau>0$ and $\tau<0$). The coincidence of this 
two branches indicate the remarkable symmetry of the crossing probability
with respect to the variable $\tau$.

In Fig.~\ref{fig1}b) we plot the crossing probability by crosses (bottom)
and the magnetic  susceptibility by triangles (top) as a functions of the inverse 
temperature  $\beta$  with logarithm scale for the ordinate
axis.  In Fig.~\ref{fig1}b) we indicate the position of crossover region of $\pi_{hs}$
by horizontal solid line on a level $D \exp(-c\tau_{0})$.
 For the magnetic susceptibility we mark the region with  critical behavior
\begin{equation}
\label{eq1}
\chi(\beta) \sim ( 
(\beta-\beta_{c})/\beta_{c})^{-\gamma}
\end{equation}
by  horizontal dashed lines.
We see from Fig.~\ref{fig1}b) that tails of the crossing probability
directly correspond to the critical region of the magnetic susceptibility.
In this critical region the correlation length $\xi$ is smaller than 
the sample size  $\xi<L$. As the temperature approaches to the critical point, the 
correlation length reaches the sample size. At that point  the magnetic susceptibility 
on the finite lattice deviates from the critical behavior Eq.~(\ref{eq1})
and becomes smooth -- see the  region over the top dashed horizontal line
in the Fig.~\ref{fig1}b). 
At the same point the crossing probability crosses over from tails
to body -- the region over the solid horizontal line in Fig.~\ref{fig1}b)
(and the region {\it under} the horizontal line in Fig.~\ref{fig1}a)).
 At the critical point $\beta_{c}=-\ln(1-p_{c})\simeq 0.881373\dots$
both the magnetic susceptibility and the crossing probability
reach a maximum.

The detailed description of the fitting procedure
as well as numerical data for $q=1,2,3,4$ are presented below.
The main numerical
result of this paper is the proving of the formula
$D \exp( b[p-p_{c}]L^{{\nu}})$
for the tails of the crossing probability.
Therefore, we pay special attention to fitting procedures.

The paper is organized as follows:
In the second section, we describe details of the numerical simulation.
In the third section, the method for determining
 the pseudocritical point $p_{c}(L,q)$ on the finite lattice
is described.
We use  $p_{c}(L,q)$ to perform the approximation of the tails.
In section~\ref{secapp1} we approximate the
double logarithm of the crossing probability $\ln(-\ln(\pi_{hs}(p;L,q)))$ 
tails as a function of the logarithm of deviation from the critical point 
$\ln(p-p_{c}(L,q))$ by the linear function
$\tilde c(L,q) + x(L,q) \ln(p-p_{c}(L,q)) $. We get 
$x(L,q) \simeq \nu(q)$ for this approximation procedure.
In section~\ref{secapp2} we describe  new fitting procedure
 using  the  scaling variable $\tau=L(p-p_{c})^{\nu}$.
Results of approximation are discussed in  Section~\ref{secres}.

\section{Details of numerical simulation}

\begin{figure}[th]
\begin{center}
\epsfig{file=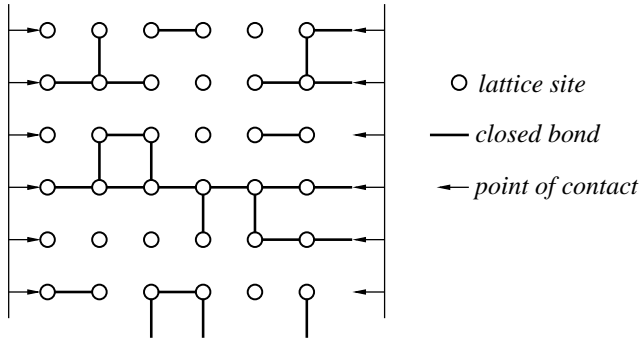,width=8.6cm,height=4.5cm}
\end{center}
\caption{
The dual lattice with the horizontally spanning cluster.}
\label{fig2}
\end{figure}

We perform the massive Monte-Carlo simulation 
on the square lattice of size $L$ to obtain the 
high-precision data for $\pi_{hs}$. We use the dual lattice shown 
in Fig.~\ref{fig2}.  On such  lattice the critical point of the 
bond percolation ($q=1$) is exactly equal $1/2$ and is not 
dependent on the lattice size~\cite{ZN}. To produce the 
pseudorandom numbers we use the R9689 random number generator
with four taps~\cite{Ziff}. We  close each bond with a probability $p$
and leave it open with  a  probability $1-p$.
Then we split the lattice into clusters of connected sites
by using the Hoshen-Kopelman algorithm~\cite{HK}. 
After that we check the percolation through this configuration.
We average the crossing probability
over $10^{7}$ random bond configurations. 
To investigate the Potts model $q=2,3,4$ we use the Wolff~\cite{Wolff}
cluster algorithm to generate a sequence of 
thermally equilibrated 
spin configurations. For each particular inverse temperature 
$\beta=\frac{1}{T}$ we flip 20000 Wolff clusters to equilibrate the system.  
For the spin model on the finite lattice
the deviation of the pseudocritical point from the 
position of the critical point on the infinite lattice is smaller
for Periodic Boundary Condition (PBC)
rather than the Open Boundary Condition (OBC)~\cite{Landau1,Landau2}.
For this reason  we use the  PBC for the Wolff algorithm.
The Monte-Carlo algorithm generates spin configurations on a
torus. For a generated spin configuration
we create a configuration of bonds. Each bond between sites with 
equal spin variable $\sigma$   is closed with the probability 
$p=1-\exp(-\beta)$ and is open with probability $1-p=\exp(-\beta)$.
Bonds between sites with different values of $\sigma$  are always
open in accordance with Fortuin-Kasteleyn rule~\cite{FK}.
  Then we split the particular spin and bond configurations
into  different clusters.
 Here we use OBC. It means that for each  generated configuration
 we cut the torus and check the crossing on the square
 with open boundary conditions.
We fix the OBC for crossing only in horizontal
direction $\pi_{hs}$ because it implies the vertical crossing is absent
and the top and bottom rows must be disjointed. But we take into
consideration the additional raw and column of bonds, as shown in 
Fig.~\ref{fig2}.
We check the percolation through an
obtained cluster configuration, 
generate new spin configuration and so on. 
We average the crossing probability over $10^{7}$
configurations for each value of the inverse temperature $\beta$.
 So the resolution of our computations is about $10^{-7}$.
By this way we perform numerical simulation and get a set of data for 
$\pi_{hs}(p;L,q)$ for the lattice sizes $L=32,48,64,80,128$ and 
$q=2,3,4$. The formal definition of $\pi_{hs}(p;L,q)$ as a sum over
different cluster configurations is described in~\cite{Vas}.

For the Potts model we use the dual lattice as we do for the percolation. 
It means, that we take into account  additional bonds
attached to the bottom row of spins.  
In the same way we take into account
 additional bonds attached to the right column of the spins.
On the  lattice with PBC
these bonds have to connect the right and the left columns.
 We cut the torus (because we use OBC for crossing probability)
 but we keep these additional bonds and take
into account  the checking of the crossing. 
 In Fig.~\ref{fig2}
 contact points are shown by arrows. The left
contact points  are attached to the left column of sites.
The right contact points are attached to  additional bonds. 
In Fig.~\ref{fig2} the bond configuration with the horizontally 
spanning cluster is shown. 
Then we check the percolation through the obtained cluster configuration.
After that we flip three Wolff clusters, check spanning
for a new spin configuration and so on.

\section{
Determination of the pseudocritical point on the finite lattice}
\label{secpc}

We investigate the crossing probability
as a function of deviation from the critical point.
Therefore, we perform the preliminary approximations 
to obtain the critical points for the finite samples.
Namely, we: 
\begin{itemize}
\item obtain the pseudocritical point and the shape of the 
      central part of the crossing probability,
\item determine the shape of the tails of the crossing probability 
\item combine together the information for body and tails,
      and reconstruct the total shape of the crossing probability. 
\end{itemize}

We need to recall that we consider the crossing probability as a function
of the variable $p=1-\exp(-\beta)$ (probability of a bond to be closed).
It is easy to understand that we must take the pseudocritical point
on the finite lattice $p_{c}(L,q)$ as the reference point. 
The crossing probability is a symmetric
function of the variable $\Delta p=(p-p_{c}(L,q))$.
This fact implies that the high temperature tail 
$p_{c}-\Delta p$ ($\Delta p>0$) and the low temperature
tail $p_{c}+\Delta p$
 coincide $\pi_{hs}(p_{c}-\Delta p)=\pi_{hs}(p_{c}+\Delta p)$. 
For the bond percolation on the dual lattice the 
position of the percolation point  does not depend
on the lattice size $p_{c}(L,q=1)=0.5$~\cite{ZN}.

For the determination of $p_{c}(L,q)$ for the Ising and the Potts model we 
use the following procedure.
 We can assume~\cite{Vas} that in the  
\mbox{ region $-6<\ln(\pi_{hs}(p;L,q))$} the  fitting formula 
is true
\begin{equation}
\label{eq2}
F(p;L,q)= A(q,L)\exp(  -  \left[ B(L,q)(p-p_{c}(L,q))
\right]^{\zeta(L,q)}).
\end{equation}
Therefore we fit the logarithm of the crossing probability
$\ln(\pi_{hs})$ by the function $f_{1}(p;L,q)=\ln(F(p;L,q))$
namely
\begin{equation}
\label{eq3}
f_{1}(p;L,q)= a(L,q))  -  \left[ B(L,q)(p-p_{c}(L,q))
\right]^{\zeta(L,q)},
\end{equation}
where $a(L,q)=\ln(A(L,q))$.
\begin{figure*}[t]
\mbox{
\epsfig{file=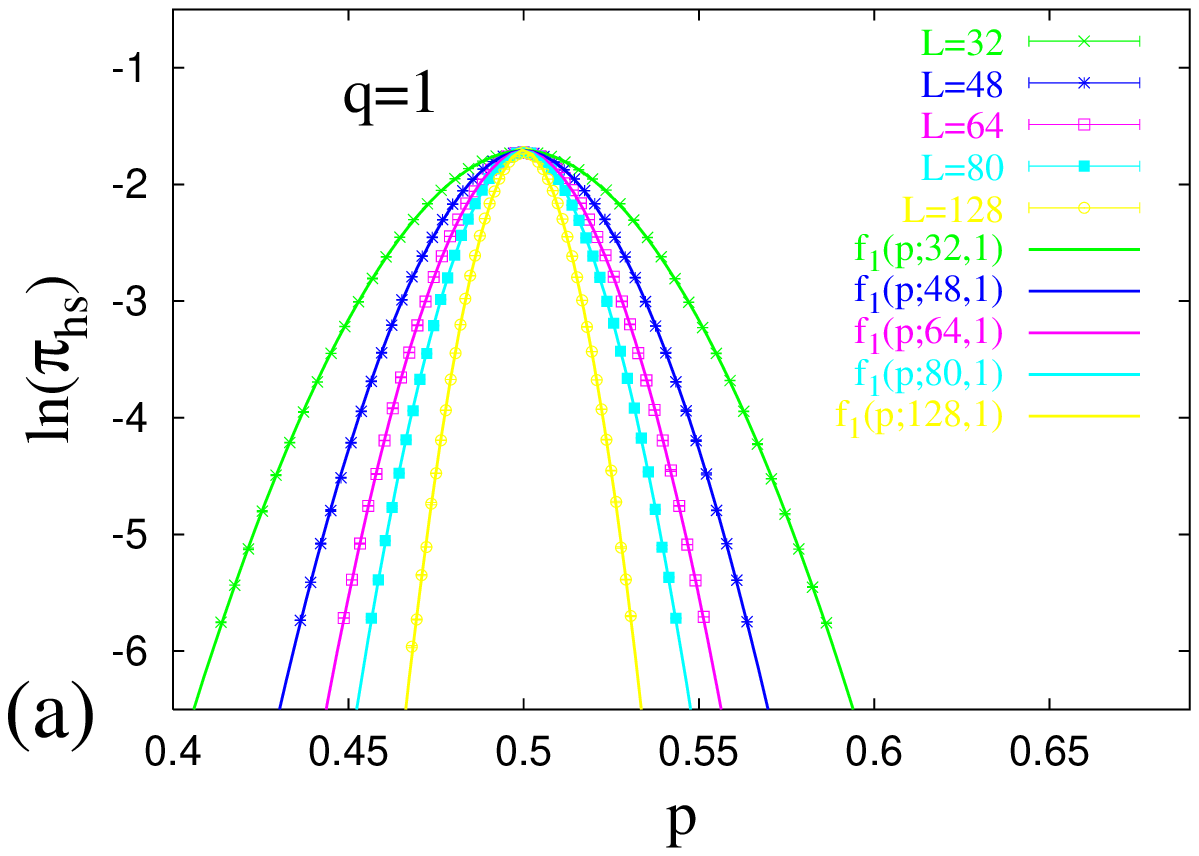,width=8.6cm,height=7cm}
\epsfig{file=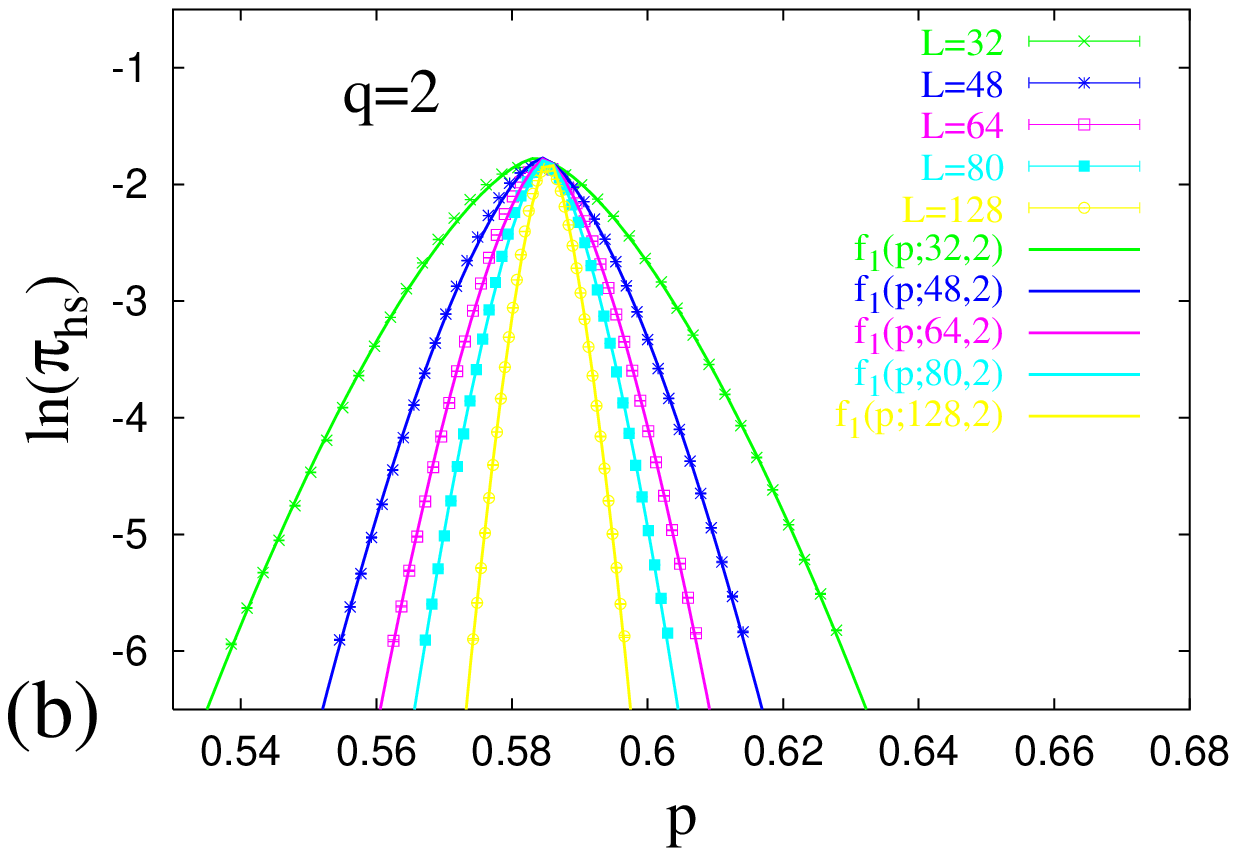,width=8.6cm,height=7cm}}
\mbox{
\epsfig{file=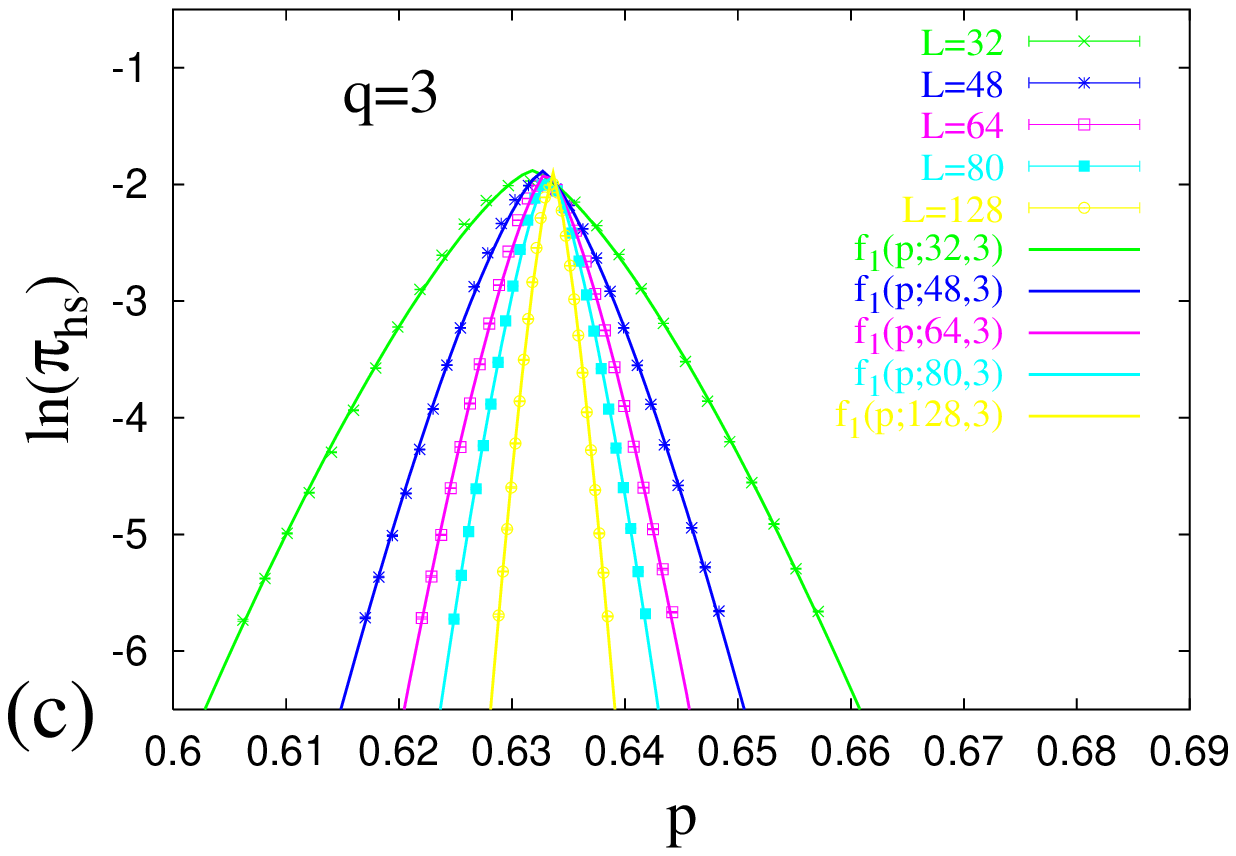,width=8.6cm,height=7cm}
\epsfig{file=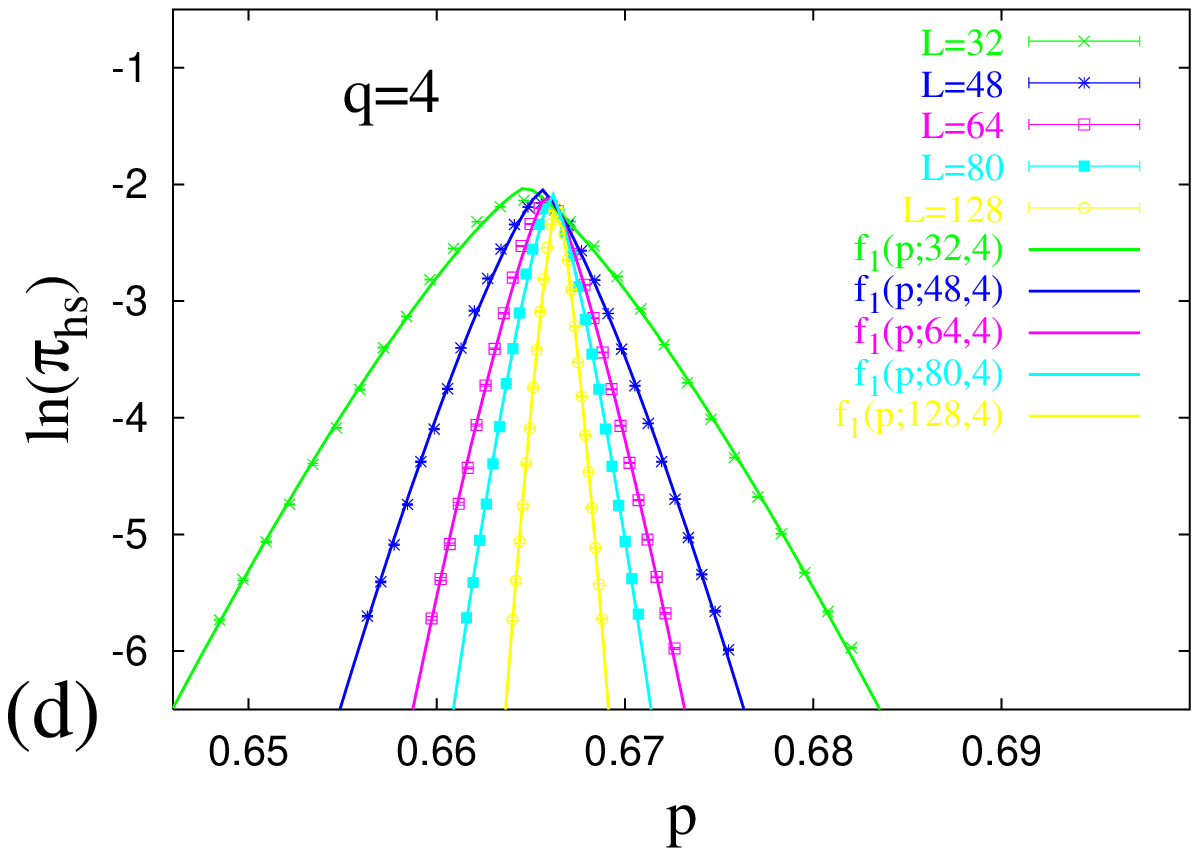,width=8.6cm,height=7cm}}
\caption{
Approximation of the logarithm of the crossing
probabilities $\ln(\pi_{hs})$ as a function of $p$
by the function
$f_{1}(p;L,q)=a(L,q)-B(L,q)(p-p_{c}(L,q))^{-\zeta(L,q)}$
for: a)  percolation $q=1$, b) Ising model $q=2$, c) Potts model $q=3$,
d) Potts model $q=4$.
}
\label{fig3}
\end{figure*}
We plot the data for $\ln(\pi_{hs}(p;L,q))$ as a function of $p$
in Fig.~\ref{fig3}a), Fig.~\ref{fig3}b), Fig.~\ref{fig3}c), Fig.~\ref{fig3}d)
for $q=1,2,3,4$ respectively. The errorbars in these figures are
 about the symbol size. It seems, that behavior of $\ln(\pi_{hs}(p;L,q))$
near $p_{c}$ is parabolic. 
Results of the approximation are plotted
in the same figures by lines. We shall see that 
there is a good agreement between the numerical data and the results of 
the approximation. But we see deviation at the point $p_{c}$ 
 especially for $q=3,4$. In the vicinity of
$p_{c}$ real graphs are more smooth than fitting functions.
Finally for each pair of numbers $(L,q)$ we obtain  four fitting 
parameters $A(L,q),\;B(L,q),\;p_{c}(L,q),\;\zeta(L,q)$. 
Here $A(L,q)$ defines the crossing probability in the critical 
point, $B(L,q)$ is a scaling variable,  $p_{c}(L,q)$ 
is the position of the pseudocritical point on the lattice $L$,
and $\zeta(L,q)$ is a scaling index.

In Table~\ref{tab1} we collect data for the logarithm 
of critical amplitude $a(L,q)=\ln(A(L,q))$. 
The fitting parameter $a(L,q)$ defines a vertical shift of  graphs
in Fig.~\ref{fig3}a)-\ref{fig3}d) from the zero level.
 \begin{table}[h]
\caption{Results of the approximation for the 
fitting parameter $a(L,q)=\ln(A(L,q))$.}
\begin{tabular}{|c|c|c|c|c|}
\hline 
 $L$ & $q=1$ & $q=2$ & $q=3$ & $q=4$ \\
\hline
\hline                                                                         
 32 &  -1.708(4)   & -1.773(14)   & -1.881(23)   & -2.022(27)  \\               
 \hline                                                                         
 48 &  -1.710(4)   & -1.773(16)   & -1.886(27)   & -2.043(30)  \\               
 \hline                                                                         
 64 &  -1.711(4)   & -1.776(16)   & -1.894(27)   & -2.055(32)  \\               
 \hline                                                                         
 80 &  -1.711(5)   & -1.777(17)   & -1.897(27)   & -2.078(29)  \\               
 \hline                                                                         
 128 &  -1.705(8)   & -1.777(18)   & -1.898(29)   & -2.089(31)  \\   
\hline
\end{tabular}
\label{tab1}
\end{table} 
In Table~\ref{tab2} the data for the scaling variable $B(L,q)$
are shown.
We approximate the data for $B(L,q)$ by the function
$ b^{*}(L,q)L^{u(L,q)}$.
\begin{figure}[ht]
\epsfig{file=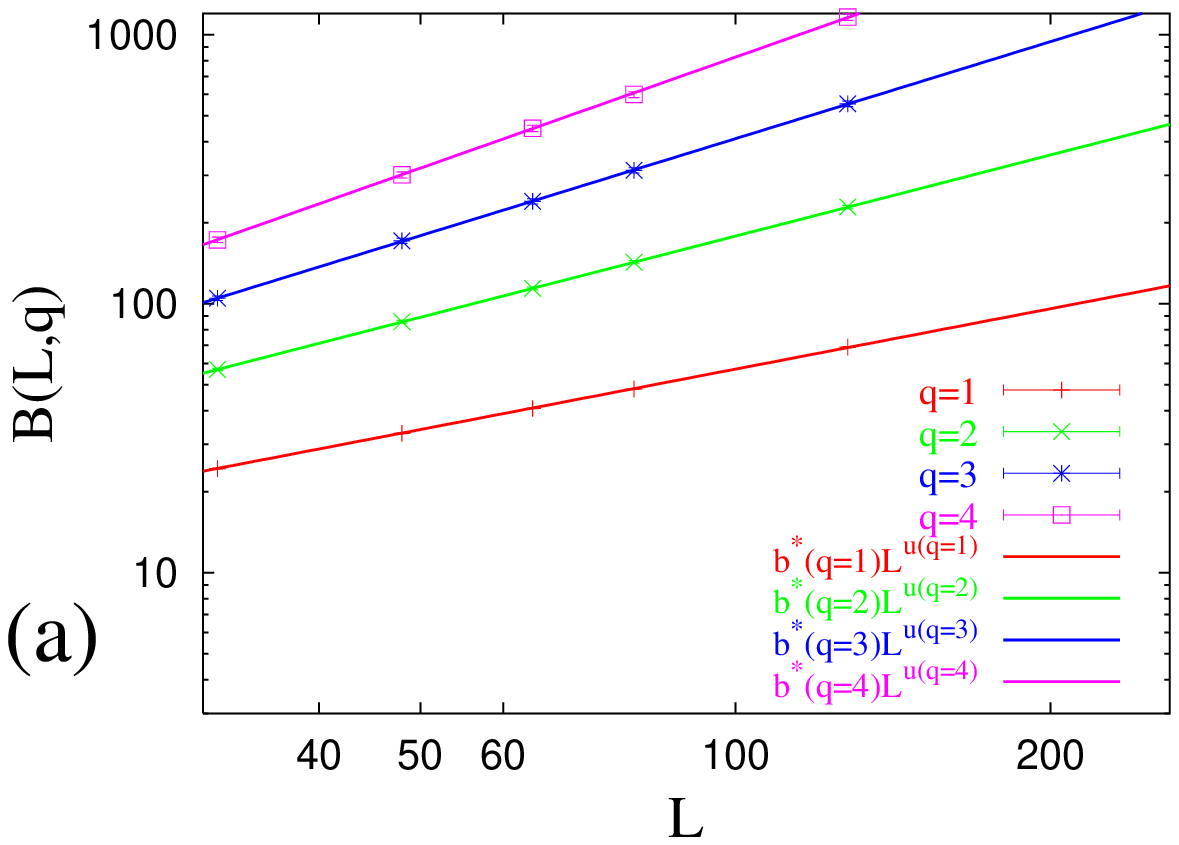,width=8.6cm,height=7cm}
\epsfig{file=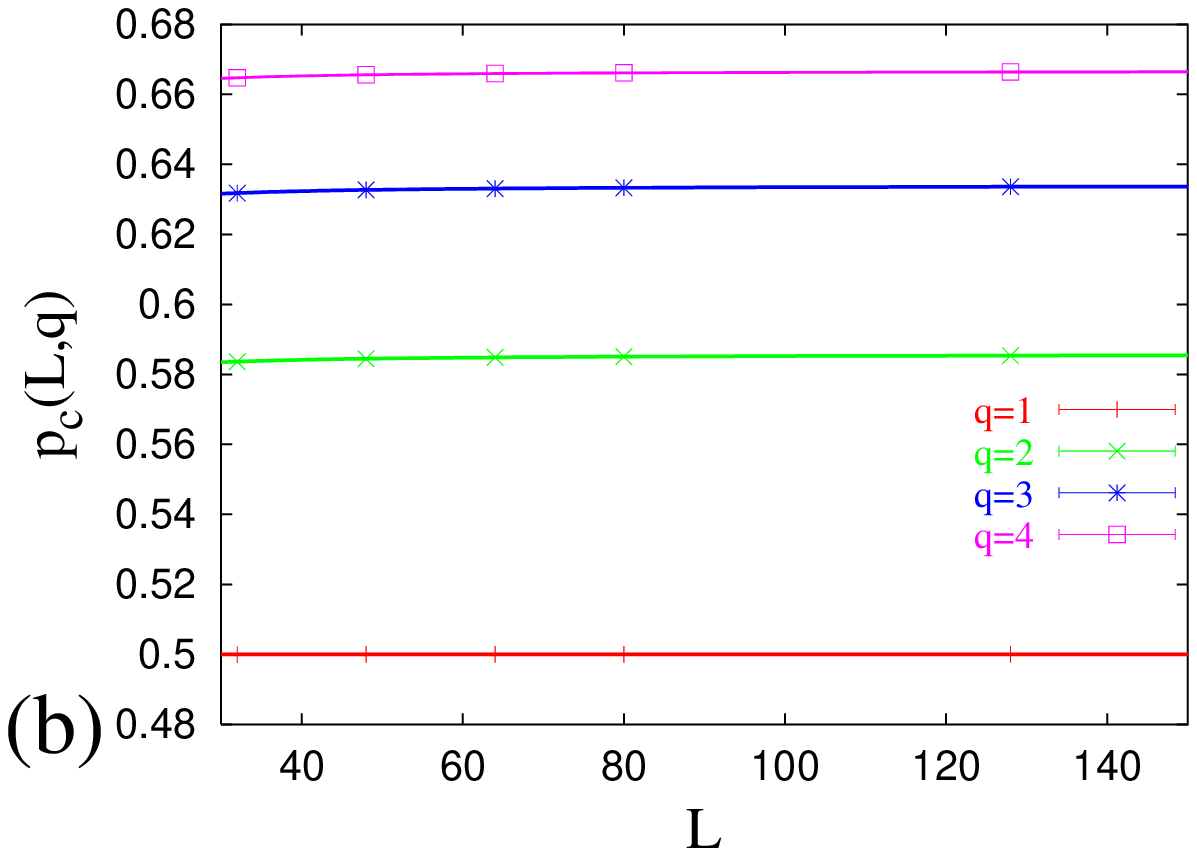,width=8.6cm,height=7cm}
\caption{
a) The fitting parameter $ B(L,q)$ 
as a function of $L$. Results of the approximation by
functions $ b^{*}(L,q)L^{u(L,q)}$ (see Table~\ref{tab2}) are plotted 
by lines;
b) Position of the pseudocritical 
point on the finite lattice $p_{c}(L,q)$ as a function of the lattice size 
$L$ for $q=1,2,3,4$ and approximation by the pow law 
$p_{c}(L,q)=p_{c}(\infty,q)+ dp(q)L^{v(q)}$ for $q=2,3,4$.
Line $0.5$ is  also added.
}
\label{fig4}
\end{figure}
In Fig.~\ref{fig4}a) the results of approximation are plotted by lines.
 Values of the fitting parameters $ b^{*}(L,q)$, $u(L,b)$
as well as the inverse correlation length index $\frac{1}{\nu}$
 are placed in the three last rows of Table~\ref{tab2}.
We can see for $q=1,2,3$ that  $u(L,q)\simeq \frac{1}{\nu(q)}$.
It can be assumed that 
\begin{equation}
\label{eq4}
B(L,q)= b^{*}(L,q)L^{\frac{1}{\nu(q)}}.
\end{equation}
In the case $q=4$ the scaling index $u(q)$ is not equal 
$\frac{1}{\nu(q=4)}$.
Many critical quantities in the Potts model $q=4$
exhibit logarithmic corrections~\cite{CNS,SS,AA,CZ}.
These logarithmic corrections explain the difference
between analytical value $\frac{1}{\nu(q=4)}=1.5$
and numerical approximation for the scaling index $u=1.372(8)$.
\begin{table}[h]
\caption{Data for $B(L,q)$ and fitting parameters $ b^{*}(L,q) $,
$u(L,q)$.}
\begin{tabular}{|c|c|c|c|c|}
\hline 
 $L$ & $q=1$ & $q=2$ & $q=3$ & $q=4$ \\
\hline
\hline
 32 &  24.8(1)  &  60(1)  & 116(3)  & 195(6) \\
\hline
\hline                                                                          
 32 &  24.4(1)  &  57(1)  & 105(2)  & 173(4) \\                                 
\hline                                                                          
 48 &  33.0(1)  &  86(1)  & 171(4)  & 301(8) \\                                 
\hline                                                                          
 64 &  40.8(1)  & 114(1)  & 240(5)  & 449(12) \\                                
\hline                                                                          
 80 &  48.2(1)  & 143(2)  & 313(7)  & 599(15) \\                                
\hline                                                                          
 128 &  68.9(4)  & 229(3)  & 553(13)  & 1163(31) \\ 
 \hline
$ b^{*}(L,q)$ & 1.84(2) &1.767(9)   & 1.63(2) & 1.49(5) \\
\hline
$u(L,q)$        & 0.746(3)& 1.002(1)  &  1.198(4) & 1.372(8) \\
\hline
$\frac{1}{\nu(q)}$ & 0.75 & 1 & 1.2 & 1.5 \\
\hline
\end{tabular}
\label{tab2}
\end{table} 
The position of the critical point
 as a function of the lattice size  $L$ is shown in Fig.~\ref{fig4}b).
On the dual lattice the position of the critical point for percolation is
equal $\frac{1}{2}$ and does not depend on the lattice size.
For our computation for all points $p_{c}(L,q)$ the deviations
from $0.5$ are less than $0.0001$.
This deviation corresponds to numerical inaccuracy of our Monte-Carlo 
simulation. 
\begin{table}[h]
\caption{Results of the approximation of $p_{c}(L,q)$
by the power low $p_{c}(q)+dp(q)L^{v(q)}$. The analytical values
$p_{c}^{precise}(q)=\sqrt{q}/(\sqrt{q}+1)$ are added for comparison. 
}
\begin{tabular}{|c|c|c|c|c|}
\hline
 $q$ & $dp(q)$ & $v(q)$ & $p_{c}(q) $ & $ p_{c}^{precise}(q)$ \\  
\hline
\hline
2 & -0.160(5) & -1.25(1) & 0.58573(2) & 0.585786... \\
\hline
3 & -0.22(2) & -1.33(2) & 0.63394(2) & 0.633975.. \\
\hline
4 & -0.26(2) & -1.42(4) & 0.66666(1)  & 0.6666666.. \\
\hline
\end{tabular}
\label{tab3}
\end{table}
The data $p_{c}(L,q)$ for $q=2,3,4$ were approximated by 
the power function of the lattice size $p_{c}(L,q)\simeq 
p_{c}(q)+dp(q)L^{v(q)}$. 
Results of approximation are placed in Table~\ref{tab3}
and are also plotted by lines in Fig.~\ref{fig4}b).
We shall see that our fitting procedure determines
the critical point with accuracy up to four  digits after the decimal 
point.
Thus, we shall use obtained values $p_{c}(L,q)$
for following approximation of  tails of the 
crossing probability.
In Table~\ref{tab4} we place the results of the approximation 
for the index $\zeta(L,q)$ describing the curvature of the central part
of the crossing probability. 
\begin{table}[h]
\caption{Results of the approximation for $\zeta(L,q)$.}
\begin{tabular}{|c|c|c|c|c|}
\hline 
 $L$ & $q=1$ & $q=2$ & $q=3$ & $q=4$ \\
\hline
\hline
 32 &  1.887(6)   & 1.52(2)   & 1.38(3)   & 1.28(3)  \\                   
 \hline                                                                         
 48 &  1.883(7)   & 1.52(2)   & 1.37(3)   & 1.27(3)  \\                   
 \hline                                                                         
 64 &  1.882(6)   & 1.52(2)   & 1.38(3)   & 1.28(3)  \\                   
 \hline                                                                         
 80 &  1.885(7)   & 1.52(2)   & 1.38(3)   & 1.30(3)  \\                   
 \hline                                                                         
 128 &  1.86(1)   & 1.52(2)   & 1.37(3)   & 1.28(3)  \\   
 \hline
\end{tabular}
\label{tab4}
\end{table} 

\section{The shape of the the crossing probability {\normalsize 
$\pi_{hs}$} tails: direct approach. }

\label{secapp1}

\begin{figure*}[t]
\mbox{\epsfig{file=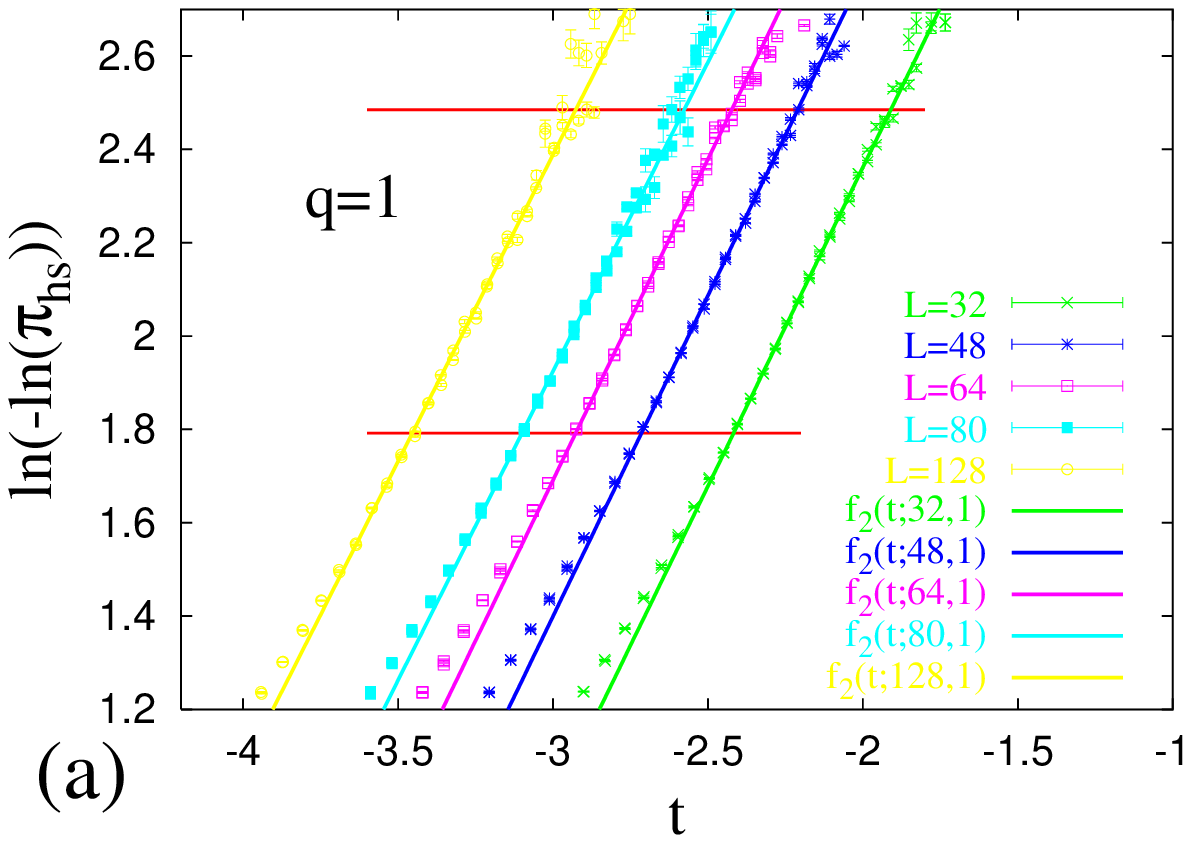,width=8.6cm,height=7cm}
\epsfig{file=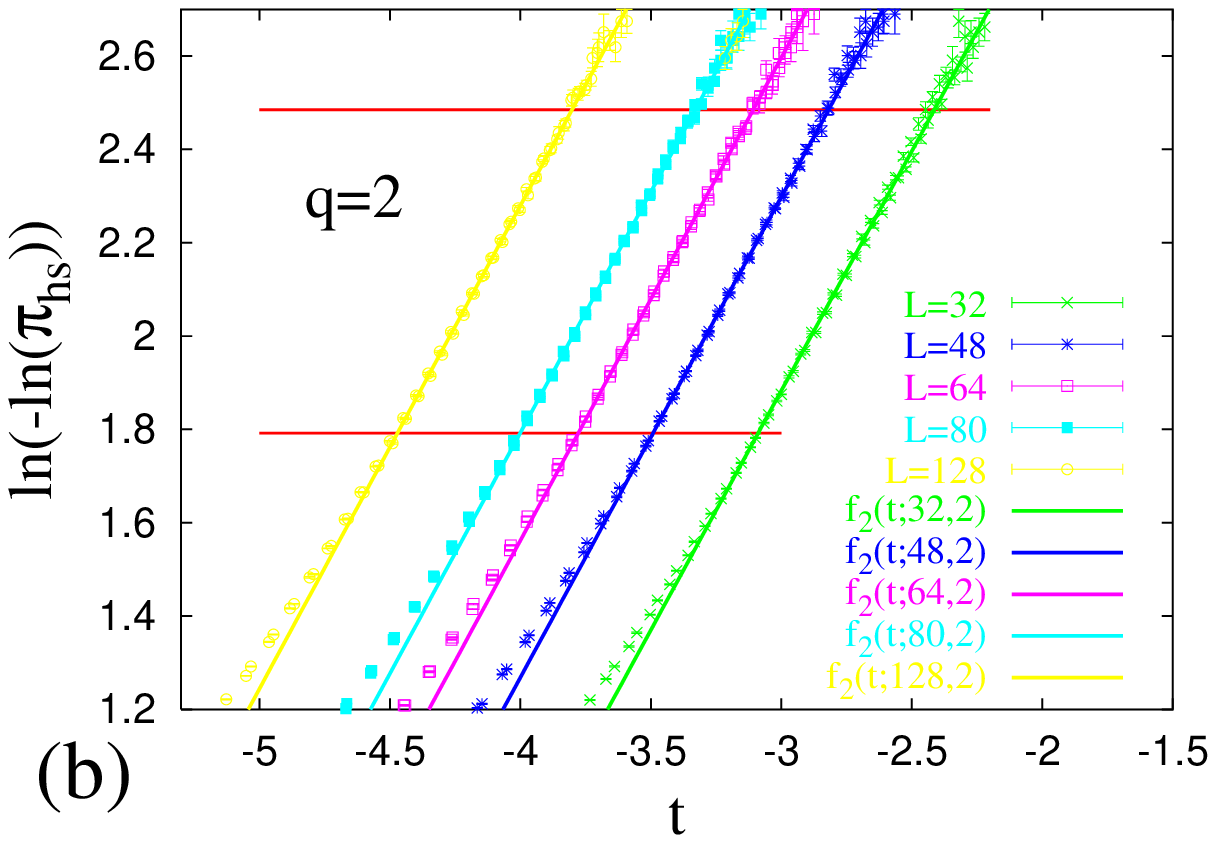,width=8.6cm,height=7cm}}
\mbox{\epsfig{file=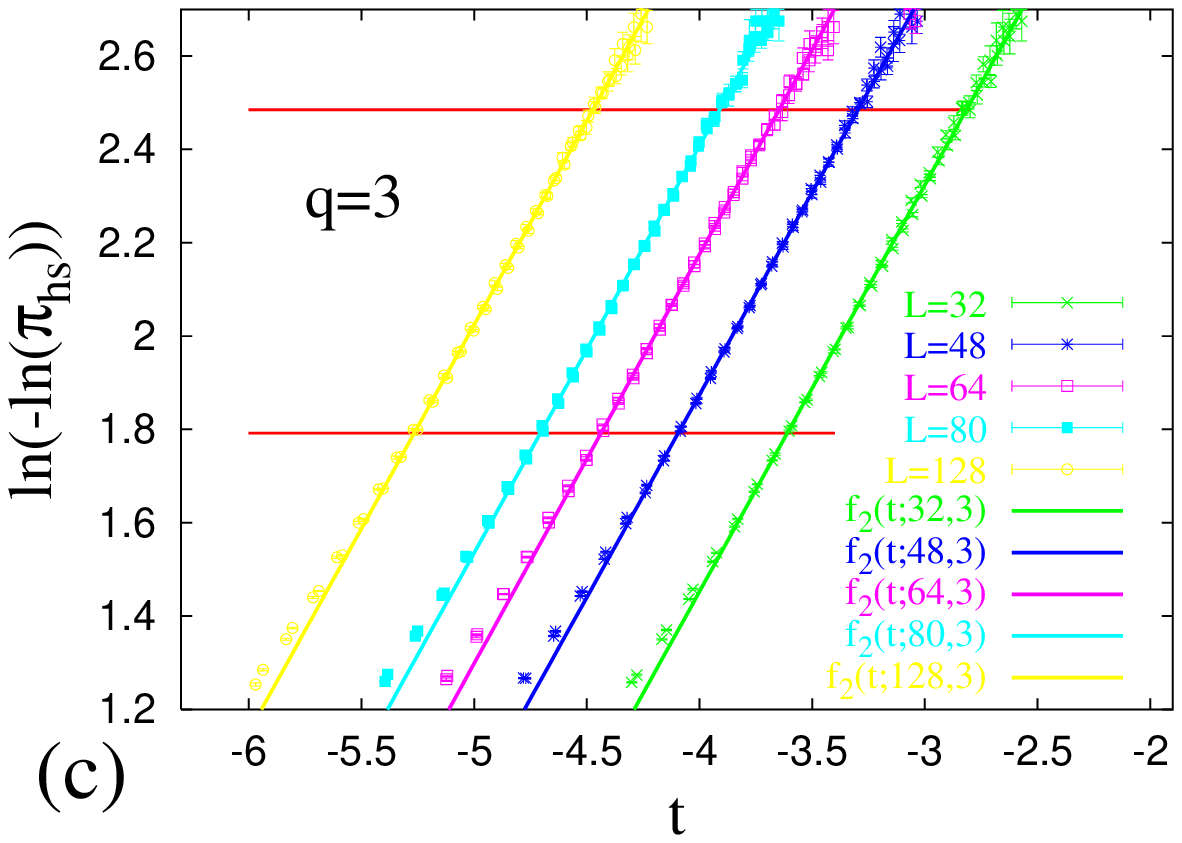,width=8.6cm,height=7cm}
\epsfig{file=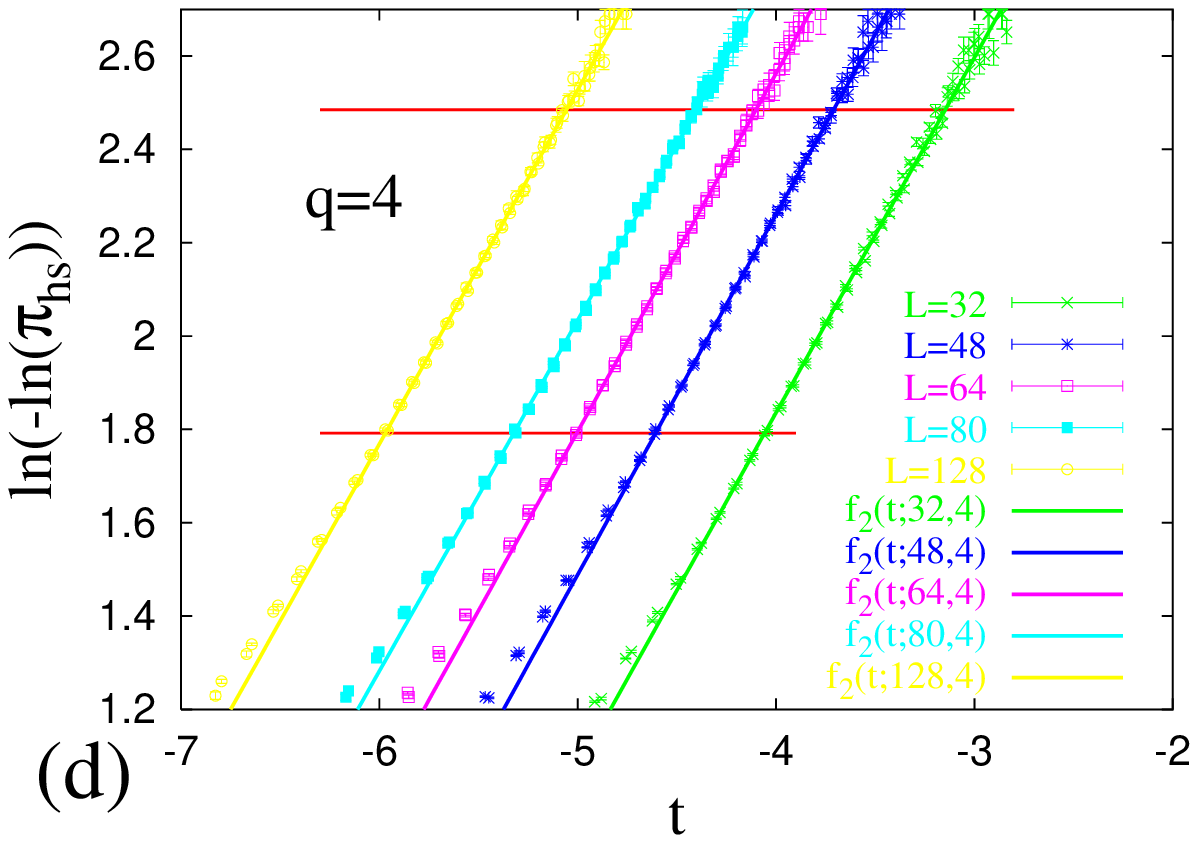,width=8.6cm,height=7cm}}
\caption{
Approximation of the double logarithm of the crossing
probabilities $\ln(-\ln(\pi_{hs}(p;L,q)))$ by 
the function   $f_{2}(t;L,q)=\tilde c(L,q)+x(L,q)t$
of the logarithm of the distance to the critical point
$t=\ln(|p-p_{c}(L,q)|)$
for: a)  percolation $q=1$, b) Ising model $q=2$, c) Potts model $q=3$,
d) Potts model $q=4$.
The range of the approximation is shown by  horizontal lines.
}
\label{fig5}
\end{figure*}

Let us check the shape of the  crossing probability tails.
The double logarithm of the crossing probability 
$\ln(-\ln(\pi_{hs}(L,q)))$ 
are plotted as   functions of the variable  $t=\ln(|p-p_{c}(L,q)|)$ 
in Fig.~\ref{fig5}a), Fig.~\ref{fig5}b), Fig.~\ref{fig5}c), 
Fig.~\ref{fig5}d) for $q=1,2,3,4$ respectively. 
Points for high-temperature and 
low-temperature  tails 
lay on the same lines. This yields that we shall prove one more time  
our fitting procedure.

We expect that the crossing probability tails 
are described by the formula
\begin{equation}
\label{eq5}
\pi_{hs}(p;L,q)=D(L,q) \exp\left(-C(L,q) (p-p_{c}(L,q))^{x(L,q)} \right).
\end{equation}
In the interval $\ln(6) \le \ln(-\ln(\pi_{hs}(L,q)))\le \ln(12)$
 the tails of the crossing probability look like straight 
lines.

The absence of prefactor before the exponent in Eq.~(\ref{eq5})
was argued in Ref.\cite{NZ2} for the case of wrapping in the horizontal 
direction in terms of transfer-matrix. We can not
directly apply the same arguments in our case, because we 
consider another function -- the probability of crossing only in the 
horizontal direction. The presence
of prefactor must cause the deviation from the linear dependence in 
Fig.~\ref{fig5} in accordance with Eq.~(\ref{eq6})
\begin{equation}
\label{eq6}
\begin{array}{l}
\ln\left[-\ln(\pi_{hs}) \right] =
\ln\left[
-d(L,q)+C(L,q)\exp(t x(L,q)) \right] \simeq \\
\simeq \ln(C(L,q))+ x(L,q)t-\frac{ d(L,q)) }{ 
C(L,q)\exp(t x(L,q)) }.
\end{array}
\end{equation}
Here $d(L,q)=\ln(D(L,q))$ and
$t=\ln(|p-p_{c}(L,q)|)$ is a logarithm of deviation from the critical 
point. But deviation due to prefactor $D(L,q) \ne 1$
 exponentially decreases as $t$ growth so it is possible
to avoid the deviation from the linear dependence
by appropriate choice of an interval of approximation.
 In the region of approximation points in Fig.~\ref{fig5}  lie on  
lines with
good accuracy.
We obtain the set of $p_{c}(L,q)$ for $L=32,\dots,128$ and $q=2,3,4$ in 
Section~\ref{secpc}. Using  this data let us approximate 
the double logarithm of tails $\ln(-\ln(\pi_{hs}))$ as 
a  function of $t$ by formula~(\ref{eq7})
\begin{equation}
\label{eq7}
f_{2}(t;L,q)= \tilde c(L,q)+ x(L,q)t.
\end{equation}
Combining~(\ref{eq5}) and~(\ref{eq7}),
we obtain $\tilde c(L,q)=\ln(C(L,q))$. 
  The resolution of our computations  $10^{-7}$ is about $16$ units 
in $(-\ln)$ scale  and is about $2.7$ units in $(\ln (-\ln))$ scale.
We use the points in the interval $\ln(6)< 
\ln(-\ln(\pi_{hs}))<\ln(12)$,
$\ln(6)\simeq 1.79$, $\ln(12) \simeq 2.48$
for this approximation. This interval is indicated in 
Fig.~\ref{fig5}a)-Fig.~\ref{fig5}d) by   horizontal lines.
 In this figures the  slope lines represent  results of 
 approximation for $x(L,q)$. 

\begin{figure}[ht]
\epsfig{file=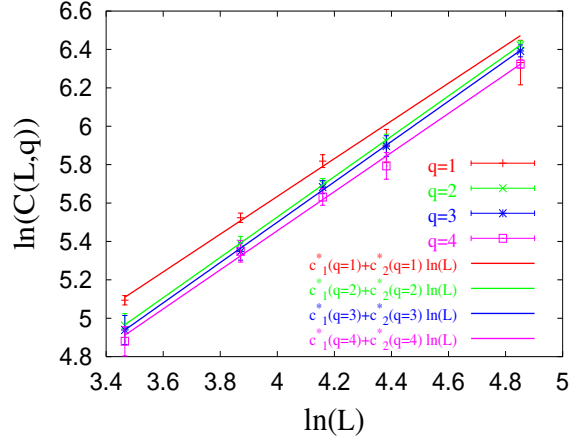,width=8.0cm,height=6cm}
\caption{
The fitting parameter $ \tilde c(L,q)=\ln(C(L,q))$ 
(the vertical shift of the approximation lines for tails)
as a function of $\ln(L)$ 
for $q=1,2,3,4$. Results of approximation
$c_{1}^{*}(q)+c_{2}^{*}(q)\ln(L)$ are shown by lines. 
}
\label{fig6}
\end{figure}

In the region $\ln(-\ln( \pi_{hs})) < \ln(6)$ 
the crossing probability obeys  another scaling formula. 
In the region  
 $\ln(-\ln(\pi_{hs})) > \ln(12) $ the numerical inaccuracy  becomes 
large.
There are  only a few numbers of  ``hits'' in horizontally spanning  
clusters for $\ln(-\ln(\pi_{hs})) > \ln(12) $.
Results of the  numerical approximation for $x(L,q)$ and $\tilde c(L,q)$
are presented in Table~\ref{tab5} and Table~\ref{tab6} respectively.
\begin{table}[h]
\caption{The slope of  lines: results of the approximation for $x(L,q)$,
the analytical values   $\nu(q)$ are presented for comparison.}
\begin{tabular}{|c|c|c|c|c|}
\hline 
 $L$ & $q=1$ & $q=2$ & $q=3$ & $q=4$ \\
\hline
\hline
$\nu$   & $4/3\simeq 1.3333$ & $1$ & $5/6\simeq 0.83333$  & $2/3\simeq 
0.6666$ \\  
\hline
\hline                                                                         
 32 &  1.366(11)   & 1.026(23)   & 0.87(2)   & 0.761(21)  \\                  
 \hline                                                                         
 48 &  1.375(10)   & 1.030(12)   & 0.868(12)   & 0.772(14)  \\                  
 \hline                                                                         
 64 &  1.376(12)   & 1.035(8)   & 0.877(8)   & 0.767(10)  \\                    
 \hline                                                                         
 80 &  1.32(3)   & 1.032(10)   & 0.872(13)   & 0.752(14)  \\                  
 \hline                                                                         
 128 &  1.31(4)   & 1.036(5)   & 0.874(6)   & 0.759(4)  \\      
 \hline
\end{tabular}
\label{tab5}
\end{table} 
As we can see   $x(L,q) \simeq \nu(q)$. Exception is the case $q=4$.
For $q=4$ we obtain $x(L,q=4)\simeq 3/4$ 
instead of $\nu(q=4)=2/3$. This deviations can be explained
by the logarithmic corrections.  Some deviations  exceeding 
the approximation errors can be explained by
the choice of the approximation region. As it can be seen
from Fig.~\ref{fig5}a)--Fig.~\ref{fig5}d) 
decreasing  of the bottom approximation bound $\ln(6)$
causes  decreasing of the slop of the approximation line.
From all said above we can conclude that the tails of the 
crossing probability behaves like $\exp(-(p-p_{c})^{\nu})$.
\begin{table}[h]
\caption{Shift of the lines: results of the approximation 
$\ln(C(L,q))=\tilde 
c(L,q)=c_{1}^{*}+c_{2}^{*}\ln(L)$.}
\begin{tabular}{|c|c|c|c|c|}
\hline 
 $L$ & $q=1$ & $q=2$ & $q=3$ & $q=4$ \\
\hline
\hline                                                                          
 32 &  5.09(2)  & 4.96(6)  & 4.940(8)  & 4.88(8) \\                      
\hline                                                                          
 48 &  5.52(2)  & 5.39(4)  & 5.35(5)  & 5.35(6) \\                      
\hline                                                                          
 64 &  5.82(3)  & 5.70(3)  & 5.68(3)  & 5.63(4) \\                      
\hline                                                                          
 80 &  5.9(8)  & 5.92(4)  & 5.90(5)  & 5.79(7) \\                      
\hline                                                                          
 128 &  6.33(12)  & 6.42(2)  & 6.39(3)  & 6.32(2) \\  
\hline
\end{tabular}
\label{tab6}
\end{table} 
We fit the  parameter $\tilde c(L,q)$ by the expression 
$  c^{*}_{1}+  c_{2}^{*}\ln(L)$ and show results in 
Table~\ref{tab7}.
In Fig.~\ref{fig6} the fitting parameter $\tilde c(L,q)$
is plotted as a function of $L$ for $q=1,2,3,4$.
Results of approximation $c_{1}^{*}+c_{2}^{*}\ln(L)$ for $q=1,2,3,4$
are added as well.
We may assume that the parameter $ c_{2}^{*}$ for $q=1,2,3,4$ is
equal 1 within accuracy of the approximation. 
In accordance with scaling theory for the system of size $L$ the
 deviation from the critical point  is 
described by expression $L(p-p_{c})^{\nu}$.
\begin{table}[h]
\caption{Results of the approximation of $\tilde c(L,q)$ by 
the function $c_{1}^{*}(q)+ c_{2}^{*}(q) 
\ln(L)$.}
\begin{tabular}{|c|c|c|c|c|}
\hline
 $q$ & $1$ & $2$ & $3$ & $4$ \\
\hline
\hline
$ c_{1}^{*}$ & 1.7(2) & 1.32(12) & 1.30(6) & 1.38(1) \\
\hline
$ c_{2}^{*}$ & 0.98(6) & 1.05(2) &  1.05(2) & 1.02(2) \\
\hline
\end{tabular}
\label{tab7}
\end{table}
Therefore we may assume that 
\begin{equation}
\label{eq8}
\pi_{hs}(p;L,q)\simeq D(L,q)\exp \left(-c(q) L \left[ 
p-p_{c}(L,q)\right]^{\nu(q)} 
\right).
\end{equation}
In Fig.~\ref{fig1}a) we show the crossover from 
the parabolic like functional dependence
 in the vicinity of the critical point to the tails. 
Namely, we plot the data $-\ln(\pi_{hs})$ for $q=2$,
$L=128$ as a function of the variable $\tau=L(p-p_{c}(L,q))^{\nu}$. 
We shall call the function under consideration
 in the vicinity of the critical point 
a "body" of the crossing probability.  
Procedure  of approximation for this figure is described below.

\section{New fitting procedure  }
\label{secapp2}

From the above said we can make the following conclusions.
There are two scaling regions: the first one
is in the vicinity of  the critical point, and the second is the tails.
Analyzing the Eqs.~(\ref{eq2}),~(\ref{eq4}) and~(\ref{eq8})
we can conclude that the distance from the critical point
$(p-p_{c}(L,q))$ and the lattice size $L$ occur
in a formula only as combination $L/\xi \sim L(p-p_{c}(L,q))^{\nu(q)}$
in accordance with scaling theory.
Let us introduce a scaling variable
\begin{equation}
\label{eq9}
\tau=L(p-p_{c}(L,q))^{\nu(q)}.
\end{equation}
If we plot the negative logarithm of the crossing probability
$-\ln(\pi_{hs})$ as a function of the scaling variable $\tau$ then we
expect the power dependence in the vicinity of zero 
and linear dependence 
for tails, as we can see in  Fig.~\ref{fig1}a).
We can use the scaling variable $\tau=L(p-p_{c}(L,q))^{\nu(q)}$
to fit the crossing probability taking into account the 
finite size scaling. 
Let us describe a new fitting procedure.

We may assume that we obtain the position of the critical point
on the finite lattice $p_{c}(L,q)$  as a result of the  previous fit.
Thus, we can use only three free fitting parameters 
and fix the value $p_{c}(L,q)$.
 Substituting Eq.~(\ref{eq9}) for $\tau$ in Eq.~(\ref{eq3})
and~Eq.~(\ref{eq5}), we obtain  the fitting formulas
for the body and the tails of the crossing probability
\begin{equation}
\label{eq10}
-\ln(\pi_{hs}) \simeq f_{3}(\tau;L,q)= -a(L,q)+b(L,q)\tau^{z(L,q)},
\end{equation}
\begin{equation}
\label{eq11}
-\ln(\pi_{hs}) \simeq  f_{4}(\tau;L,q)=-d(L,q)+c(L,q)\tau.
\end{equation}
Here $\tau=L(p-p_{c}(L,q))^{\nu(q)}$.
If our assumptions are correct then
 fitting parameters $a,d,c,z$ do not depend on the lattice size $L$.

The numerical value of the variable $p_{c}(L,q)$ is used
for computation of $\tau$ in accordance with~Eq.~(\ref{eq9}). 
Comparing  new fitting procedures Eqs.~(\ref{eq10}),~(\ref{eq11}) and  
previous fitting procedures~Eqs.~(\ref{eq3}),~(\ref{eq5})
we can obtain relations between  new fitting parameters 
$b(L,q)$, $z(L,q)$, $c(L,q)$ and parameters 
$ b^{*}(L,q)$, $\zeta(L,q)$, $C(L,q)$
\begin{equation}
\label{eq12}
b(L,q)= b^{*}(L,q)^{ \zeta(q)},
\end{equation}
\begin{equation}
\label{eq13}
\zeta(L,q)=z(L,q)\nu(L,q),
\end{equation}
\begin{equation}
\label{eq14}
 C(L,q)=c(L,q)L,\;  \tilde c(L,q)= \ln (C(L,q)). 
\end{equation}
The special case is $q=4$. As we can observe from the results of the 
approximations in Table~\ref{tab2} and Table~\ref{tab5}
the finite size scaling of the crossing probability
for $q=4$ does not obey the scaling low $L(p-p_{c})^{\nu(q=4)}$
(probably to logarithmic corrections to scaling). 
Therefore, instead of the analytical
value $\nu(q=4)=2/3$ we use the numerical approximation 
$x(q=4)= 3/4$ from Table~\ref{tab5}.
We can see later, that this substitution
allows us to obtain correct results for scaling.

We perform the fitting procedure in accordance with  the formula proposed 
above.
For the fit of the body of the crossing probability~Eq.~(\ref{eq10})
 the scaling region $1.6<-\ln(\pi_{hs})<6$ was used.
For the fit of the tails of the crossing 
probability~Eq.~(\ref{eq11})
we use the scaling region $6<-\ln(\pi_{hs})<12$.
\begin{figure*}[t]
\mbox{\epsfig{file=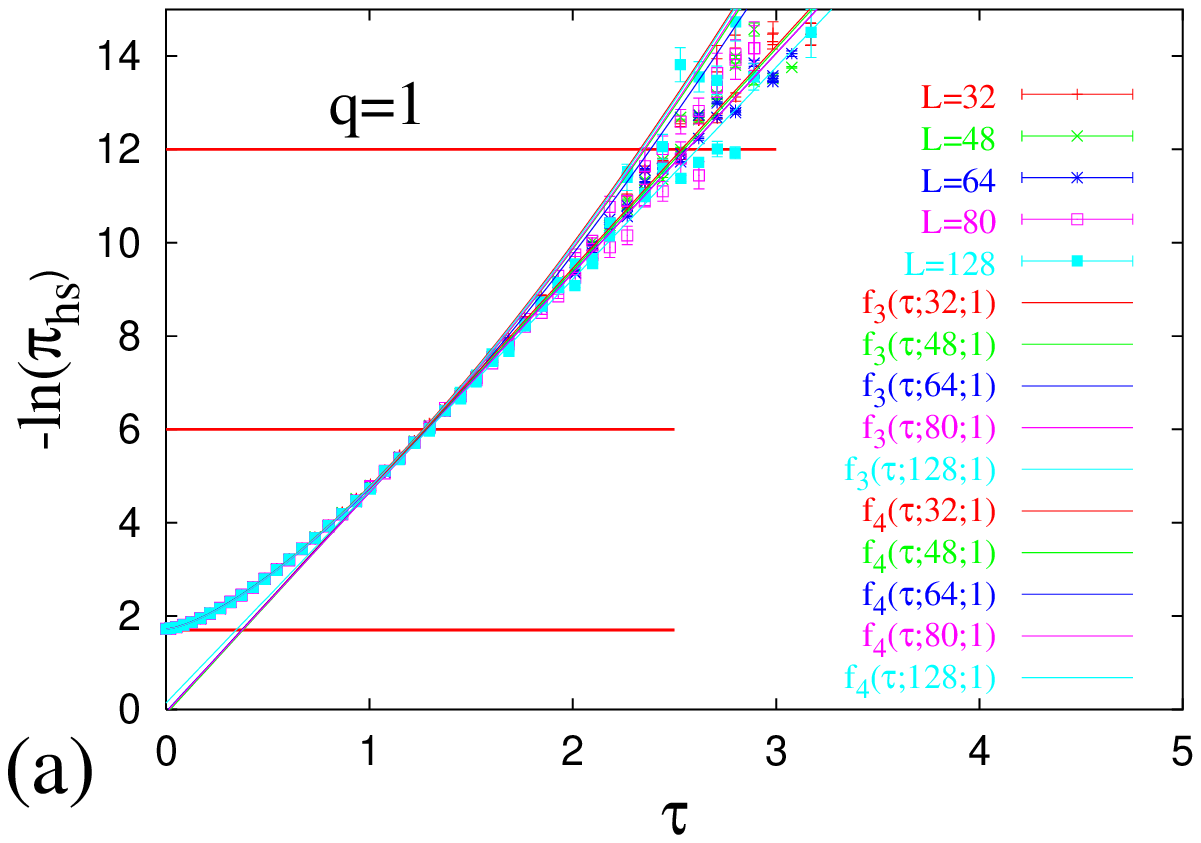,width=8.6cm,height=7cm}
\epsfig{file=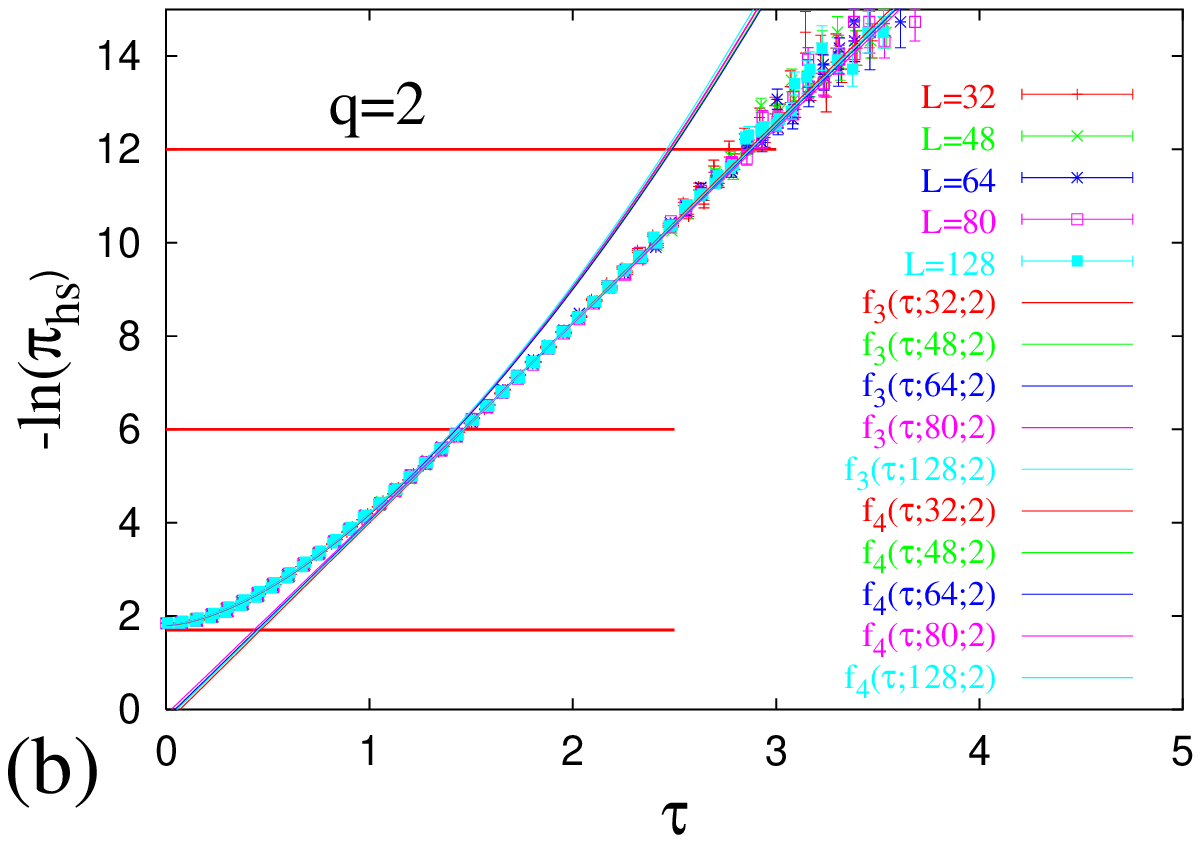,width=8.6cm,height=7cm}}
\mbox{\epsfig{file=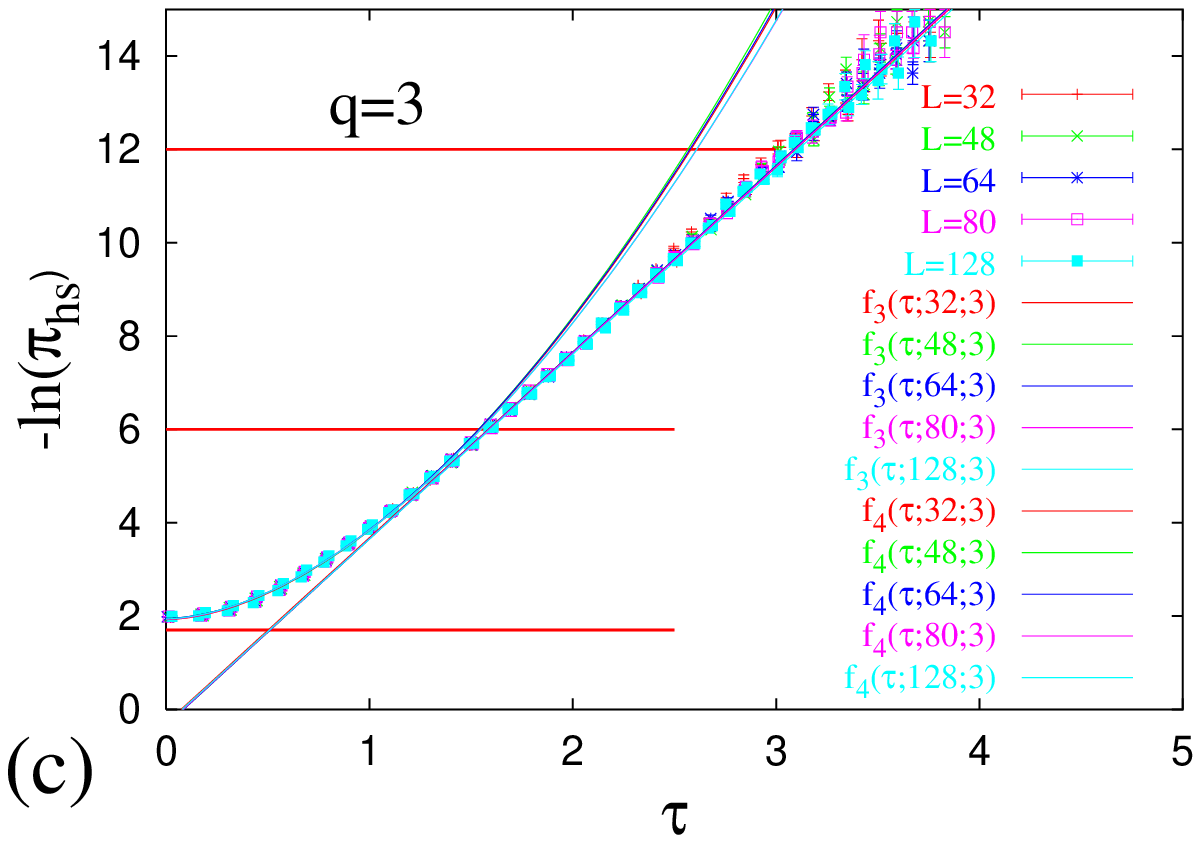,width=8.6cm,height=7cm}
\epsfig{file=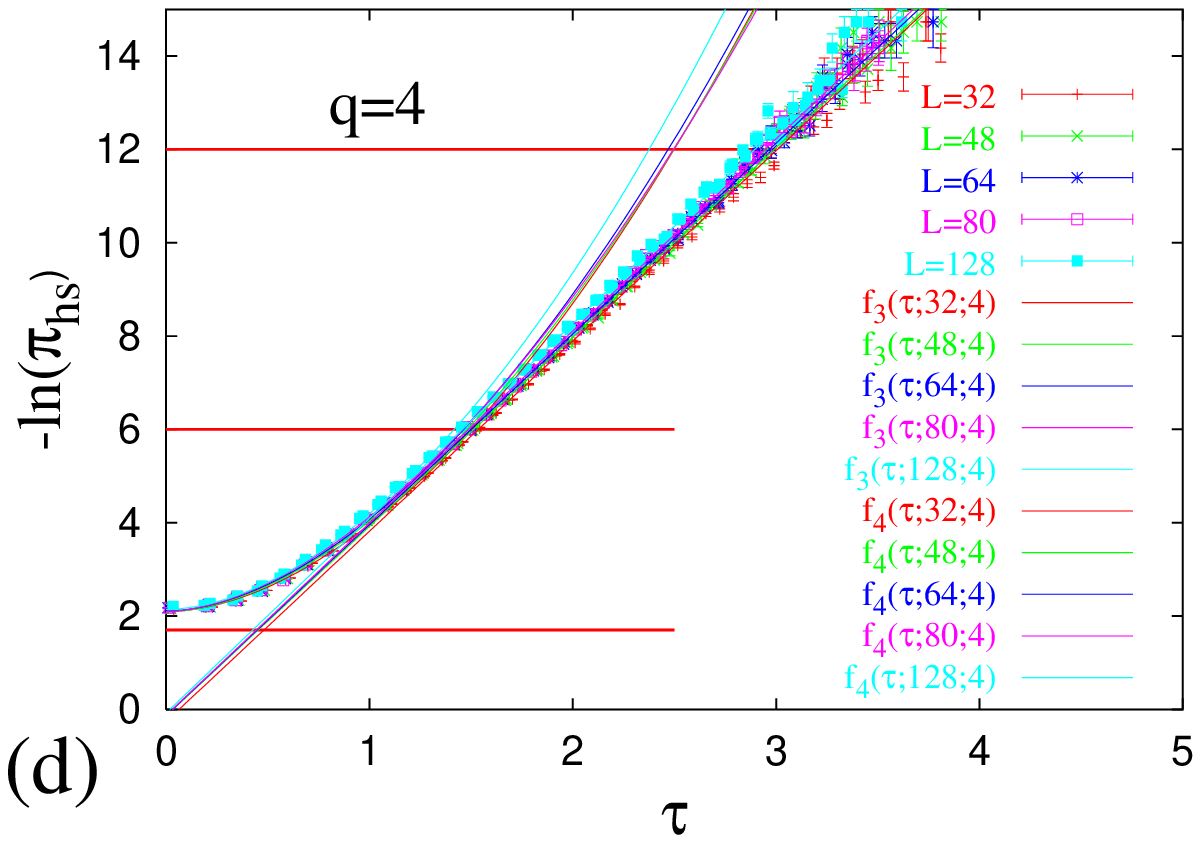,width=8.6cm,height=7cm}}
\caption{
The negative logarithm of the crossing probability  
$-\ln(\pi_{hs}(p;L,q))$ as a function of the
 the scaling variable
$\tau=L(p-p_{c}(L,q))^{\nu(q)}$
for: a)  percolation $q=1$, b) Ising model $q=2$, c) Potts model $q=3$,
d) Potts model $q=4$.
Results of the approximation of the body of the crossing probability
by the function $f_{3}(\tau;L,q)=-a(L,q)+b(L,q)\tau^{z}$
and  tails of the crossing probability
by the function $f_{4}(\tau;L,q)=-d(L,q)+c(L,q)\tau $ are added.
The ranges of approximation regions are shown by  horizontal lines. 
}
\label{fig7}
\end{figure*}
We place results of the fitting procedure for $q=1,2,3,4$ in 
Fig.~\ref{fig7}a), Fig.~\ref{fig7}b), 
Fig.~\ref{fig7}c), Fig.~\ref{fig7}d) respectively.
In these figures the fitting regions are denoted 	 
by horizontal lines.
As we expect, if we plot the data as  functions of the
scaling variable $\tau$ then 
the finite size dependence  
eliminates and points for different values of $L$
 lie on the same curves.
Results of the approximation for the fitting parameters
$b(L,q)$, $z(L,q)$, $c(L,q)$ and $d(L,q)$
are collected in Table~\ref{tab8}, Table~\ref{tab9},
Table~\ref{tab10}, Table~\ref{tab11}
respectively.

The data for $b(L,q)$ in Table~\ref{tab8}
demonstrates, that this fitting parameter does not depend
on the lattice size $L$.
All dependence on $L$ is enclosed in the expression for $\tau$.
Hence we can omit the variable $L$ in the round brackets and
consider $b(q)$ only as a function of $q$.
This fact proved the choice of the fitting procedure.
\begin{table}[h]
\caption{Results of the approximation for the 
fitting parameter $b(L,q)$.}
\begin{tabular}{|c|c|c|c|c|}
\hline 
 $L$ & $q=1$ & $q=2$ & $q=3$ & $q=4$ \\
\hline
\hline
 32 &  3.063(3)  & 2.38(2)  & 1.91(2)  & 1.87(3) \\                       
\hline                                                                          
 48 &  3.032(3)  & 2.40(2)  & 1.89(2)  & 1.95(2) \\                       
\hline                                                                          
 64 &  3.018(5)  & 2.38(2)  & 1.90(2)  & 1.97(3) \\                       
\hline                                                                          
 80 &  3.027(5)  & 2.39(2)  & 1.91(3)  & 2.02(3) \\                       
\hline                                                                          
 128 &  3.049(6)  & 2.38(2)  & 1.90(3)  & 2.05(3) \\ 
\hline
\end{tabular}
\label{tab8}
\end{table} 
The data in Table~\ref{tab9}
presents some power $z(q)$. This power 
describes the behavior of the crossing probability
in the vicinity of the critical point as a function of the scaling 
variable $\tau$ and does not depend on the lattice size. 
\begin{table}[h]
\caption{Results of the approximation for the 
fitting parameter $z(L,q)$.}
\begin{tabular}{|c|c|c|c|c|}
\hline 
 $L$ & $q=1$ & $q=2$ & $q=3$ & $q=4$ \\
\hline
\hline
 32 &  1.432(3)   & 1.61(2)   & 1.75(3)   & 1.82(3)  \\                   
 \hline                                                                         
 48 &  1.426(4)   & 1.59(2)   & 1.77(3)   & 1.78(3)  \\                   
 \hline                                                                         
 64 &  1.413(5)   & 1.59(2)   & 1.76(4)   & 1.78(4)  \\                   
 \hline                                                                         
 80 &  1.432(4)   & 1.60(3)   & 1.73(4)   & 1.73(3)  \\                   
 \hline                                                                         
 128 &  1.432(6)   & 1.61(3)   & 1.74(4)   & 1.81(5)  \\  
\hline
\end{tabular}
\label{tab9}
\end{table} 
The data for the fitting parameter $c(L,q)$
are placed in Table~\ref{tab10}. It seems, that this parameter
does not depend on  $L$.
\begin{table}[h]
\caption{Results of the approximation for the 
fitting parameter $c(L,q)$.}
\begin{tabular}{|c|c|c|c|c|}
\hline 
 $L$ & $q=1$ & $q=2$ & $q=3$ & $q=4$ \\
\hline
 \hline
 32 &  4.76(3)  & 4.29(4)  & 3.97(3)  & 4.08(4) \\                              
\hline                                                                          
 48 &  4.75(3)  & 4.25(1)  & 4.00(2)  & 4.06(2) \\                              
\hline                                                                          
 64 &  4.70(2)  & 4.23(2)  & 4.00(3)  & 4.09(2) \\                              
\hline                                                                          
 80 &  4.71(6)  & 4.19(3)  & 3.99(3)  & 4.12(2) \\                              
\hline                                                                          
 128 &  4.54(8)  & 4.26(4)  & 3.96(3)  & 4.17(2) \\     
\hline
\end{tabular}
\label{tab10}
\end{table} 
The results for $d(L,q)$ is placed in Table~\ref{tab11}. 
First of all, this parameter does not depend on the lattice
size within accuracy of the approximation.
The absolute value of parameter $d(q)$ is relatively small
so the prefactor $D(q)$ is about 1. For percolation
the order of $d(q=1)$ is less than three value of numerical inaccuracy,
thus it is possible $d(q=1)=0$.

\begin{table}[h]
\caption{Results of the approximation for the 
fitting parameter $d(L,q)$.}
\begin{tabular}{|c|c|c|c|c|}
\hline 
 $L$ & $q=1$ & $q=2$ & $q=3$ & $q=4$ \\
\hline
\hline
 32 &  -0.07(4)   & -0.30(7)   & -0.29(5)   & -0.26(7)  \\                      
 \hline                                                                         
 48 &  -0.10(4)   & -0.21(3)   & -0.35(3)   & -0.14(4)  \\                      
 \hline                                                                         
 64 &  -0.05(3)   & -0.19(4)   & -0.34(6)   & -0.13(3)  \\                      
 \hline                                                                         
 80 &  -0.06(9)   & -0.11(5)   & -0.33(6)   & -0.13(3)  \\                      
 \hline                                                                         
 128 &  0.14(12)   & -0.23(7)   & -0.31(6)   & -0.08(3)  \\   
\hline
\end{tabular}
\label{tab11}
\end{table}

\section{Discussion}
\label{secres}

Using the dual lattice (see~Fig.~\ref{fig2})
allows us to avoid finite size shift of the critical point
for the bond percolation and to 
diminish it for  spin models. The accuracy of definition 
of the critical point on the finite lattice
play the principal role
for the investigation of the tails scaling.
The high quality of our approximation is proved by
 remarkable symmetry of the crossing probability with respect
to the critical point $p_{c}$. In Fig.~\ref{fig5} and Fig.~\ref{fig7}
we can observe that the two branches $p-p_{c}>0$
and $p-p_{c}<0$ practically coincide.

The two different scaling region of the crossing probability
clearly seen in Fig.~\ref{fig1}a) can explain
the long time uncertainty about its shape.
In Ref.\cite{Wester} the scaling index for the percolation threshold
for $2d$ percolation model
was found $\zeta \simeq 1.9(1)$.
This result  coincides with our approximation of the body 
of the crossing probability for percolation $\zeta \simeq 1.864(12)$.
In more recent works~\cite{Ziff1,BW,HA,NZ1,NZ2} the
tails region for percolation was investigated
 which is described by the scaling formula  $D\exp(-c L 
(p-p_{c})^{\nu=4/3})$.
The crossover form to Gaussian-like behavior to
slope $4/3$ is observed in figures of  Ref.\cite{ONS}.
It seems, near the critical point
 the behavior of the crossing probability is parabolic.
The rounding happens in the interval 
$\tau <  0.1$.
This interval is relatively small in comparison with
regions of the body $0.1 < |\tau |< 1.5$ and tails $ 1.5< |\tau|<4  $
as it can be seen in Fig.~\ref{fig7}.

We have five fitting parameters $a(q)$, $b(q)$, $z(q)$, $c(q)$, $d(q)$
in expressions~(\ref{eq10}) and~(\ref{eq11}).
In Fig.~\ref{fig7}a)-Fig.~\ref{fig7}d) we see the crossover region 
between the body and the tails of the crossing probability. 
In this region the function $f_{3}$ touched the line $f_{4}$.
It means, that in some point $\tau_{0}$ the values of the functions are equal,
therefore 
\begin{equation}
\label{eq15}
-a(L,q)+b(q)\tau_{0}^{z(q)}=-d(L,q)+c(q)\tau_{0},
\end{equation}
and first derivatives of this functions are equal too
\begin{equation}
\label{eq16}
z(q) b(q)\tau_{0}^{z(q)-1}=c(q).
\end{equation}
Substituting $c(q)$ from~Eq.~(\ref{eq16}) in~Eq.~(\ref{eq15})
we obtain expression for $b(q)$
\begin{equation}
\label{eq18}
b(q)=\frac{d(q)-a(q)}{(z(q)-1)\tau_{0}^{z(q)}}.
\end{equation}
If the crossing probabilities at the critical points $A(q)$
(and logarithms $a(q)$) can be calculated analytically
by conformal field methods (at least for the 
percolation it is possible)~\cite{Cardy0,Cardy1,Watts}
then only four independent parameters $b(q)$, $\tau_{0}(q)$ and $z(q)$, 
$a(q)$  remain for the crossing probability.

 The main
statements for the crossing probability $\pi_{hs}$ are: 
\begin{itemize}
\item 
In accordance with scaling theory the finite size scaling of the 
crossing probabilities may be eliminated by introducing
the scaling variable $\tau=L(p-p_{c}(L,q))^{\nu(q)}$.
The crossing probability as a function of
$\tau$ does not depend on the lattice size $L$. 
\item
The body of the crossing probability scales
$\pi_{hs}(\tau)\simeq A(q)\exp(-b(q)\tau^{z(q)})$.
\item
The tails of the crossing probabilities
 scales $\pi_{hs}(\tau)\simeq D(q)\exp(-c(q)\tau)$.
The value of parameter $D(q)$ is about 1. 
\item
The finite size scaling for $q=4$ does not describe by
the analytical value of the correlation length index $\nu(q=4)=2/3$.
 We obtain some scaling index $x(q=4)$.
This index $x(q=4)\simeq 0.759(4)$ for the tails of the crossing 
probability (see Table~\ref{tab5}) or $\frac{1}{u(q)} \simeq 0.728$ 
for the body of the crossing probability (see Table~\ref{tab2}).
\end{itemize}

The author would like to thank Robert M. Ziff for helpful  discussion
and Th\'ea Bellou for reading manuscript and for her remarks.

%\clearpage

\end{document}